\documentclass[12pt]{article}

\usepackage{amsmath}
\usepackage{amssymb}
\usepackage{amsfonts}
\usepackage{amscd}
\usepackage{amsbsy}
\usepackage{amsthm}
\usepackage{latexsym}
\usepackage{graphicx}
\usepackage{slashed}
\usepackage{color}
\usepackage{cite}
\usepackage{multirow}
\usepackage{tabularx}

\usepackage[utf8]{inputenc}
\usepackage[T1]{fontenc}
\usepackage{mathrsfs}
\usepackage{subfigure}

\usepackage[hyperindex]{hyperref}
\hypersetup{colorlinks=true,linkcolor=blue,citecolor=blue,urlcolor=blue}
\renewcommand{\baselinestretch}{1.2}
\def\beq{\begin{eqnarray}}
\def\eeq{\end{eqnarray}}
\newcommand{\nn}{\nonumber}

\def\ep{\epsilon}

\def\De{\Delta}

\usepackage{float} 
\usepackage{cite}  
\usepackage{titlesec}

\usepackage{ulem} 

\titleformat*{\section}{\large\bfseries}
\titleformat*{\subsection}{\normalsize\bfseries}

\begin{document}

\begin{center}

{\large\bf
Reflection positivity in a higher-derivative model with physical bound states of ghosts}

\vskip 4mm

\textbf{Manuel  Asorey$^{a}$}\footnote{
E-mail address: \ asorey@unizar.es}, 
\quad 
\textbf{Gast{\~a}o Krein$^{b}$}\footnote{
E-mail address: \ gastao.krein@unesp.br},
\quad 
\textbf{Miguel Pardina $^{a}$}\footnote{
E-mail address: \ mpardina@unizar.es},
\\
\textbf{and \ \ Ilya L. Shapiro$^{c}$}\footnote{
E-mail address: \ ilyashapiro2003@ufjf.br}


\vskip 4mm

{\sl $^{a}$
Centro de Astropart\'{\i}culas y F\'{\i}sica de Altas Energ\'{\i}as,
Departamento de F\'{\i}sica Te\'orica.
\\
Universidad de Zaragoza, E-50009 Zaragoza, Spain}

{\sl $^{b}$ Instituto de F\'{\i}sica Te\'orica, Universidade Estadual
Paulista, Rua Dr. Bento Teobaldo Ferraz, 271 - Bloco II, 01140-070
 S\~ao Paulo, SP, Brazil}

{\sl $^{c}$ Departamento de F\'{\i}sica, ICE,
Universidade Federal de Juiz de Fora
\\ Campus Universit\'{a}rio, Juiz de Fora, 36036-900, MG, Brazil}

\end{center}

\vskip 4mm

\centerline{\textbf{Abstract}}

\begin{quotation}
\noindent
The inclusion of higher derivatives is a necessary condition for a renormalizable or superrenormalizable local theory of quantum 
gravity. On the other hand, higher derivatives lead to 
classical instabilities and a loss of unitarity at the quantum 
level. A~standard way to detect such issues is by examining the 
reflection positivity condition and the existence of a Källén–Lehmann 
spectral representation for the two-point function. We demonstrate 
that these requirements for a consistent quantum theory are satisfied 
in a theory we have recently proposed. This theory is based on a 
six-derivative scalar field action featuring a pair of complex-mass 
ghost fields that form a bound state. Our results support the interpretation that physical observables can emerge from ghost dynamics in a consistent and unitary framework.

\vskip 2mm

\noindent
\textit{Keywords:} \
Quantum gravity, Higher derivatives, Ghosts,  Bound states, Reflection positivity, K\"allén-Lehmann representation
\vskip 2mm

\noindent
\textit{MSC:} \ 
81T17,  
81T10,  
83C45  

\end{quotation}

\newpage
\section{Introduction}
\label{sec1}

Quantum gravity theories involving higher-derivative fields often 
suffer from instabilities at the classical level, which are related 
to the well-known Ostrogradski instabilities~\cite{Ostrogradski}. 
In~their quantum counterparts, these theories typically contain 
one-particle states with negative kinetic energy, known as ghosts. 
The presence of such particles can produce undesirable features, 
such as appearing at infinite times in incoming and outgoing 
states~\cite{Veltman-63}. Unfortunately, it is unavoidable to 
introduce higher-derivative terms into the action of gravity in 
quantum theory~\cite{Stelle77}, or even at the semiclassical level, 
where the gravitational field is a classical background for quantum 
matter fields~\cite{UtDW}. Because of the utmost importance of the 
subject in quantum gravity, numerous remarkable attempts have been 
made to solve the problem of ghosts using different approaches. 
Notably, starting from~\cite{Tomboulis-77,salstr}, several studies 
have explored the possibility that ghosts in renormalizable quantum 
gravity with four derivatives might acquire masses with an imaginary 
component and, hence, become unstable due to loop corrections. In 
such scenarios, the ghost mode decays in the distant future, rendering 
the $S$-matrix unitary. Alternatively, there was an interesting 
proposal~\cite{Whoisafraid} to construct a theory in which massive 
ghosts are considered alongside 
the massless and healthy (i.e., stable) graviton. However, both ideas 
can hardly be implemented in the {\it minimal} renormalizable
quantum gravity with four derivatives because in this theory there is not enough room for hosting a consistent subsector of composite of ghost fields. 

On the other hand, resolving the problem of ghosts may be simpler
in superrenormalizable quantum gravity models that incorporate 
six or more derivatives \cite{highderi}. The mass spectrum in such 
models may contain real-mass ghosts accompanied by other massive, 
healthy (normal) states. Alternatively, there might exist 
ghost states with complex-conjugate masses. In both scenarios, one 
can envision a realization of Hawking's idea~\cite{Whoisafraid}, where
he speculated that complex-conjugate ghost states might be accommodated 
in a consistent quantum theory, provided they remain off-shell and do 
not contribute to the asymptotic Hilbert space. In the 
complex-mass scenario, quantum corrections are unnecessary to observe 
the emergence of the imaginary components of the masses, as described 
in one of the pioneering papers by Salam and Strathdee~\cite{salstr}. 
Within the six-derivative (or higher) quantum gravity framework, the 
complex-mass poles in the propagator are present already at tree 
level and persist under loop corrections~\cite{Modesto-complex}.

An appealing feature of superrenormalizable theories with complex-mass 
ghosts is the possibility that such ghosts do not appear in the asymptotic 
spectrum but instead form physical bound states, an idea reminiscent of 
early proposals in the context of QCD~\cite{Nakanishi} and in modern 
QCD-inspired approaches~\cite{Roberts:2020hiw}. However, implementing such 
a mechanism in quantum gravity is substantially more difficult due to the 
nonpolynomial nature of the interactions and the presence of higher-derivative 
couplings, which make the dynamics more intricate than in typical QCD models. 
In spite of this, we recently introduced a toy model~\cite{AKPS-bounds} that 
mimics certain features of higher-derivative quantum gravity and exhibits 
a bound state formed by complex-conjugate ghost fields.  Related approaches 
exploring similar ghost-confinement mechanisms have also appeared 
in~\cite{Modesto-2023,deBrito-2023}. In the present work, we address the 
question of whether the toy model of forming bound states that was constructed 
as an example in Ref.~\cite{AKPS-bounds} is a consistent quantum field theory. 

Consistent quantum field theories are constrained by a set of
basic and fundamental principles; see, e.g., \cite{OS,GJ,Weinber-I,AsoreyIntrod} 
and also~\cite{alesh} for a recent discussion concerning higher derivative
quantum gravity. Among the most relevant principles, one can list the Wightman 
axioms in Minkowski space, and in the Euclidean formalism, the Osterwalder–Schrader 
reflection positivity condition and the existence of a Källén–Lehmann spectral 
representation for the two-point function. Together with certain additional regularity 
conditions, these requirements ensure the construction of the physical Hilbert space 
and the quantum fields from the Euclidean Schwinger functions via 
the Osterwalder–Schrader theorem~\cite{OS}. The physical significance 
of these conditions has been thoroughly discussed in the aforementioned 
references, in particular, in the books~\cite{OS,Weinber-I}. 
Reflection positivity guarantees the positivity of the inner product 
in the reconstructed Hilbert space, while the Källén–Lehmann representation,
applicable to both Wightman and Schwinger two-point functions, 
encodes unitarity through the positivity of the spectral density. 
Analyticity, in turn, embodies causality by enforcing locality and 
relativistic propagation. Taken together, these conditions guarantee
unitarity, causality, and the probabilistic interpretation of the theory 
in accordance with the principles of quantum mechanics. In the present 
work, we consider the Källén-Lehmann representation for the bound state 
and the Osterwalder–Schrader condition of reflection positivity in the 
corresponding quantum theory. 

The paper is organized as follows. The next section, Sec.~\ref{sec2}, 
is a technical part of the introduction, as we discuss the link between 
reflection positivity and the Källén-Lehmann representation for the 
Euclidean two-point function. In Sec.~\ref{sec3} we analyze the
reflection positivity condition in the model developed in
Ref.~\cite{AKPS-bounds}. Sec.~\ref{sec4} discusses the 
Källén-Lehmann representation. Finally, in Sec.~\ref{secConc}
we draw our conclusions.

\section{Reflection positivity from Källén-Lehmann representation}
\label{sec2}

Let us start by analyzing the existing relation  between reflection positivity and the positivity of the
spectral representation. In particular, we shall show that  if a two-point Schwinger correlation
function $S_2(x-y)$ satisfies the Källén-Lehmann representation, it automatically verifies reflection positivity. 
Reflection positivity implies the
positivity of the product in the Hilbert space \cite{alesh}
\begin{equation}
\langle \theta f, S_2 f \rangle
= \int d^4x \: d^4y \, (\theta f)(x)\, S_2(x-y)\, f(y) \geq 0 \;,
\label{RPfromKLdef}
\end{equation}
for any continuous function $f\in C_0(\mathbb{R}^4_+)$ with compact support in the positive time half-space $(y_0 > 0)$ of $\mathbb{R}^4$, where $\theta$ is the time-reversal transformation defined by $(\theta f)(x_0,\mathbf{x})=f^*(-x_0,\mathbf{x})$. Assuming a Källén-Lehmann representation for the two-point function
\begin{equation}
S_2(p^2) = \int_0^\infty ds \, \frac{\rho(s)}{p^2+s}\;,
\qquad \rho(s)\ge 0 \; ,
\label{RPfromKLdefRP}
\end{equation}
which, after a Fourier transform, reads
\begin{equation}
S_2(x-y) = \int_0^\infty ds \; \rho(s)\, \Delta_s(x-y)\; ,
\qquad
\Delta_s(x) = \int \frac{d^4p}{(2\pi)^4}\,
\frac{e^{ip\cdot x}}{p^2+s} \; \label{GbDelta}
\end{equation}
in a position space representation.
Thus, the reflection positivity definition \eqref{RPfromKLdef}
factorizes as
\beq
\langle \theta f, S_2 f \rangle
= \int_0^\infty ds \; \rho(s)\, I_s
\label{thetaGb}
\eeq
where
\beq
I_s = \int d^4x\, d^4y \, (\theta f)(x)\, \Delta_s(x-y)\, f(y) \; . \label{Is}
\eeq
Since $f(y)=f(y_0,\mathbf{y})$ has support only at the  half space $y_0>0$,
which means it is only non-vanishing for $y_0>0$. (here and in what follows, this means the function
is defined to be non-zero in the given region)
, and
$(\theta f)(x_0,\mathbf{x}) = f^*(-x_0,\mathbf{x})$ only for $x_0<0$,
we set $x_0=-t'$ and $y_0=t$ with $t,t'\ge 0$, such that
\begin{equation}
I_s = \int_0^\infty dt \int_0^\infty dt'
\int d^3x\, d^3y \; f^*(t',\mathbf{x})\,
\Delta_s((-t',\mathbf{x})-(t,\mathbf{y}))\, f(t,\mathbf{y}) \; . \label{RPfromKLIs}
\end{equation}
After a Fourier transformation of the propagator kernel, we have
\begin{align}
\Delta_s((-t',\mathbf{x})-(t,\mathbf{y}))
&= \int \frac{d^3\mathbf{p}}{(2\pi)^3} e^{i\mathbf{p}\cdot(\mathbf{x}-\mathbf{y})}
\int \frac{dp_0}{2\pi}\; \,
\frac{e^{-ip_0(t+t')}}{p_0^2+\omega^2}, \nonumber\\
&= \int \frac{d^3\mathbf{p}}{(2\pi)^3} \,
e^{i\mathbf{p}\cdot(\mathbf{x}-\mathbf{y})}\frac{1}{2\omega}\, e^{-\omega(t+t')}\;,
\qquad \omega = \sqrt{\mathbf{p}^2+s} \; , \label{RPfromKLstep1}
\end{align}
where clearly $\omega>0$. Inserting \eqref{RPfromKLstep1} back to \eqref{RPfromKLIs} leads to
\begin{align}
I_s &= \int \frac{d^3\mathbf{p}}{(2\pi)^3}\frac{1}{2\omega}
\int_0^\infty dt \int_0^\infty dt' \,
e^{-\omega(t+t')}
\int d^3x\, d^3y \; f^*(t',\mathbf{x}) e^{i\mathbf{p}\cdot\mathbf{x}}
f(t,\mathbf{y}) e^{-i\mathbf{p}\cdot\mathbf{y}} \nonumber\\\nonumber\\
&= \int \frac{d^3\mathbf{p}}{(2\pi)^3}\frac{1}{2\omega}
\left(\int_0^\infty dt' \, e^{-\omega t'}
\int d^3x \, f^*(t',\mathbf{x}) e^{i\mathbf{p}\cdot\mathbf{x}}\right)
\left(\int_0^\infty dt \, e^{-\omega t}
\int d^3y \, f(t,\mathbf{y}) e^{-i\mathbf{p}\cdot\mathbf{y}}\right)\nonumber \\\nonumber\\
&= \int \frac{d^3\mathbf{p}}{(2\pi)^3}\frac{1}{2\omega}
\left| \int_0^\infty dt \, e^{-\omega t}
\int d^3x \, f(t,\mathbf{x}) e^{-i\mathbf{p}\cdot\mathbf{x}} \right|^2 \geq 0  \,. 
\end{align}
Since we have factorized the dot product into a manifestly positive
norm, and the Källén-Lehmann hypothesis requires $\rho(s)\ge 0$, we
have
\begin{equation}
\langle \theta f, S_2 f \rangle
= \int_0^\infty ds \; \rho(s)\, I_s\; \geq 0 .
\end{equation}
This means that the Källén-Lehmann representation with positive
spectral density implies Osterwalder-Schrader reflection
positivity. It should be remarked that the two-point correlation function
satisfies the reflection positivity condition \eqref{RPfromKLdef}, even 
after the removal of possible UV divergences, whenever they
are associated with local counterterms. 
The reason is that subtracting a polynomial term in $p^2$ from the two-point function contributes
as derivatives of a Dirac's delta function in position space. Thus,
this operation may add only contact terms that contribute solely at coincident points $x=y$, 
which are outside the support $\mathbb{R}^4_+$ of the $f$ functions. Therefore, reflection 
positivity is also preserved  by the renormalized  two-point correlation functions. 

\section{Reflection positivity in the bound state
model of complex ghosts}
\label{sec3}

Let us now consider the reflection positivity condition in the model
with a bound state formed by a pair of complex-mass ghost-like elementary
quantum states~\cite{AKPS-bounds}. The starting point of the model is a 
six-derivative scalar action with one healthy massless particle and 
two particles with complex-conjugate masses. By using Gaussian integration
in the free theory, one can find an equivalent action with three two-derivative 
scalar fields. Ignoring the massless part and introducing an interaction term, 
we arrive at the action
\beq
S_{\rm gh}
\,=\, \int d^4x \Big\{
\frac{i}{2} \varphi_1\big(-\partial^2 + m^2\big) \varphi_1
- \frac{i}{2} \varphi_2 \big(-\partial^2 + m^{*2}\big)
\varphi_2
- U(\varphi_1,\varphi_2)\Big\}\,,
\label{act2gh}
\eeq
where we choose the interaction of the form
\beq
U(\varphi_1,\varphi_2)
\,=\,
\frac{1}{4} \,\lambda_{12} \,\varphi^2_1 \varphi^2_2 \, . \label{U12}
\eeq
Here $\lambda_{12}>0$ is the coupling constant corresponding to
the interaction of the two fields. Let us now explore the composite operator 
made of the product $O_{\varphi_1\varphi_2}(x)$ of two ghosts with complex-conjugate masses
\beq
O_{\varphi_1\varphi_2}(x) = \varphi_1(x) \varphi_2(x) \, . \label{O12}
\eeq
The respective correlation function $C(x,y)$ defines the nature of a possible
bound state,
\begin{align}
C(x,y) &= C(x-y) = \langle O_{\varphi_1\varphi_2}(x) \,
O_{\varphi_1\varphi_2}(y) \rangle \nn \\[0.25cm]
&= \frac{1}{Z_{\rm gh}} \int \mathcal{D}\varphi_1 \mathcal{D}\varphi_2 \,
O_{\varphi_1\varphi_2}(x) \,
O_{\varphi_1\varphi_2}(y) \; e^{-S_{\rm gh}}\, .
\label{C12}
\end{align}

Using the first order two-point function $G_B(p)$
\beq
G_B(p)
\,= \,\int \frac{d^4k}{(2\pi)^4} \, D_{\varphi_1}(p-k)
\, D_{\varphi_2}(k) \, . \label{bubble0}
\eeq
with the elementary propagators
\beq
D_{\varphi_1}(p)
= \frac{i}{p^2 + m^2} \hspace{1.0cm}
\text{and}\hspace{1.0cm}
D_{\varphi_2}(p) = \frac{-\,i\,\,}{p^2 + m^{*2}}\,,
\label{Dphi12}
\eeq
we arrive at the momentum
representation of the resumed two-point correlation function (\ref{C12}) in the form
\begin{align}
C(p)
\,=\, G_B(p) \, \sum^\infty_{n=0} \big[\lambda_{12}\,G_B(p)\big]^n
\,\,=\,\, \frac{G_B(p)}{1 - \lambda_{12}\,G_B(p)}\,. \label{GB-pole}
\end{align}
Thus, the fundamental features of (\ref{C12}) depend on the
expression for the first order two-point function (\ref{bubble0}). 
This integral can be written in explicit form
\beq
G_B(p) \,=\,
\int \frac{d^4k}{(2\pi)^4}\,
\frac{1}{\big[(p-k)^2+m^2\big]\,\big[k^2+{m^*}^2\big]}\,. \label{bubble}
\eeq
Our purpose is to explore whether the correlation functions of the composite operator 
satisfy  Osterwalder--Schrader  reflection positivity. In the next section, we will derive 
an explicit, renormalized Källén–Lehmann representation for \eqref{bubble}, which, 
by the results of the previous section, establishes its reflection positivity. Here, 
however, we provide a direct and independent proof that does not rely on the Källén–Lehmann representation. Consider
\begin{align}
\De_m(k) &= \frac{1}{k^2+m^2}= \frac{1}{k_0^2+\mathbf k^2+m^2}
  \;,\;\;\;\;\;\;\;\;
  \De_m(x)=\int \frac{d^4k}{(2\pi)^4}e^{ik\cdot x}\Delta_m(k) \; ,
\end{align}
the free propagator in momentum space, and 
its Fourier transform to the coordinate space. Since $G_B(p)$ is a convolution of two free propagators in momentum space, its
corresponding Fourier transform to position space $G_B(x)$ is the pointwise product
and thus we need to apply the reflection positivity test \eqref{RPfromKLdef} to the function
\begin{align}
 G_B(x-y) = \Delta_m(x-y) \Delta_{m^*}(x-y)  \; .
 \label{PositionSpaceBubble}
\end{align}
Using the well-known Fourier integral
\begin{equation}
 \int\frac{dk_0}{2\pi}\,\,\frac{e^{ik_0 x_0}}{k_0^2+\mathbf k^2+m^2}
  \;=\;\frac{e^{- \omega_{\mathbf k}|x_0|}}{2\,\omega_{\mathbf k}}\; ,
  \qquad \omega_{\mathbf k}=\sqrt{\mathbf k^2+m^2} \; ,
\end{equation}
the Euclidean free propagator then has the representation
\begin{align}
 \Delta_m(x) \;=\; \int \frac{d^3\mathbf{k}}{(2\pi)^3}\,\frac{1}{2\omega_{\mathbf{k}}}\,
e^{i\mathbf{k}\cdot \mathbf{x}}\,e^{-\omega_{\mathbf{k}}|x_0|} \; .
\end{align}
Reflection positivity requires the positivity of the product in Hilbert space
\begin{align}
\langle \theta f, G_B f\rangle
= \int d^4x\,d^4y\; (\theta f)(x) \: G_B(x-y)\, f(y)\;\geq 0 \; . \label{defRP}
\end{align}
Since $f$ has support in $x_0>0$, the reflected function $\theta f$
has support only in $x_0<0$. Thus, in the integral \eqref{defRP}, it
is always satisfied that $x_0<0$, $y_0>0$.
Therefore
\[
|x_0-y_0|=-(x_0-y_0)=-x_0+y_0=|x_0|+y_0 \; ,
\]
because for $x_0<0$ we have $|x_0|=-x_0$. Now, expanding the bilinear form of reflection positivity \eqref{defRP} and inserting $G_B(x-y)$ explicitly \eqref{PositionSpaceBubble}, we have
\begin{align}
\langle \theta f, G_B f\rangle
&= \int \frac{d^3\mathbf{k}}{(2\pi)^3}\frac{1}{2\omega_{\mathbf{k}}}
\int \frac{d^3\mathbf{q}}{(2\pi)^3}\frac{1}{2\omega^*_{\mathbf{q}}}
\nonumber
\\
&
\times
\quad\int\limits_{x_0<0} d^4x \int\limits_{y_0>0} d^4y\; f^*(-x_0,\mathbf{x})\, f(y)\,
e^{i(\mathbf{k}+\mathbf{q})\cdot(\mathbf{x}-\mathbf{y})}\,
e^{-(\omega_{\mathbf{k}}+\omega^*_{\mathbf{q}})(|x_0|+y_0)} .
\label{RPstep1}
\end{align}
Now let us change the  variables $u_0=-x_0>0$
and $u=(u_0,\mathbf{x})$ with $\mathbf{u}=\mathbf{x}$. Then clearly
$f^*(-x_0,\mathbf{x})=f^*(u)$ and $|x_0|=u_0$,
so the bilinear form \eqref{RPstep1} becomes
\begin{align}
\langle \theta f, G_B f\rangle
&= \int \frac{d^3\mathbf{k}}{(2\pi)^3}\frac{1}{2\omega_{\mathbf{k}}}
\int \frac{d^3\mathbf{q}}{(2\pi)^3}\frac{1}{2\omega^*_{\mathbf{q}}}
\nn
\\
&
\times \quad \int\limits_{u_0>0} d^4u \int\limits_{y_0>0} d^4y\; f^*(u)\, f(y)\,
e^{i(\mathbf{k}+\mathbf{q})
\cdot\mathbf{u}}\,e^{-i(\mathbf{k}+\mathbf{q})\cdot\mathbf{y}}\,
e^{-(\omega_{\mathbf{k}}+\omega^*_{\mathbf{q}})(u_0+y_0)} .
\end{align}
Let us introduce the auxiliary function
\begin{align}
F_{\mathbf{k},\mathbf{q}}
\,=\, \int\limits_{y_0>0} d^4y\; f(y)\;
e^{-(\omega_{\mathbf{k}}
+\omega^*_{\mathbf{q}})y_0}\; e^{-i(\mathbf{k}+\mathbf{q})
\cdot \mathbf{y}} \; \label{Fkq}
\end{align}
and its  complex conjugate
\beq
F_{\mathbf{k},\mathbf{q}}^*
= \int_{u_0>0} d^4u\; f^*(u)\;
e^{-(\omega^*_{\mathbf{k}}+\omega_{\mathbf{q}})u_0}
\; e^{i(\mathbf{k}+\mathbf{q})\cdot \mathbf{u}}\,. \label{Fstar}
\eeq
Hence, using the modulus
$F_{\mathbf{k},\mathbf{q}}^* F_{\mathbf{k},\mathbf{q}}
= |F_{\mathbf{k},\mathbf{q}}|^2$, the bilinear form can be
expressed as
\begin{align}
\langle \theta f, G_B f\rangle
= \int \frac{d^3\mathbf{k}}{(2\pi)^3}\frac{1}{2\omega_{\mathbf{k}}}
\int \frac{d^3\mathbf{q}}{(2\pi)^3}\frac{1}{2\omega^*_{\mathbf{q}}}\; \big|F_{\mathbf{k},\mathbf{q}}\big|^2 \;\geq 0 \; .
\end{align}
Since we have been able to write the reflection positivity bilinear form 
as an integral of absolute squares with a positive measure, the first order 
two-point function $G_B(x-y)$ satisfies the reflection positivity property. 
Note that in the derivation, it is crucial that the two masses are complex 
conjugates (to end up with conjugate frequency factors) and have positive 
real parts so that the Fourier time-integrals yield decaying exponentials. 
One might be worried about the fact that the first order two-point function 
integral~\eqref{bubble} is divergent at large momentum. Since this UV
divergence is logarithmic, the simplest way to renormalize
it is via a momentum subtraction scheme, as was done in~\cite{AKPS-bounds}. 

Let us now define the first order renormalized two-point function
as
\begin{align}
 G'_B(p) = G_B(p) -  G_B(p_0) \; ,
\end{align}
where the subtraction is done at $p_0$ and therefore $G_B(p_0)$
is a constant independent of $p$. In position coordinate space, this renormalized
 two-point function reads
\begin{align}
    G'_B(x) = G_B(x) - G_B(p_0)\, \delta^{(4)}(x) \; . \end{align}
It is easy to see that the subtracted first order two-point function $G'_B(x)$ also satisfies
reflection positivity, since by the definition \eqref{defRP} we
explicitly have
\begin{align}
\langle \theta f, G'_B f\rangle
= \langle \theta f, G_B f\rangle
- G_B(p_0) \int d^4x \, d^4y \; (\theta f)(x)\, \delta^{(4)}(x-y)\, f(y) \; .
\end{align}
The delta function reduces the last integral to
\[
\int d^4x \; (\theta f)(x) f(x) \; ,
\]
but $(\theta f)(x)=f^*(-x_0,\mathbf{x})$ has support only where $x_0<0$,
while $f(x)$ has support only where $x_0>0$. Since the supports are disjoint, the delta integral contribution vanishes; thus,
\[
\langle \theta f, G'_B f\rangle = \langle \theta f, G_B f\rangle \; ,
\]
and so reflection positivity is preserved for the renormalized first order two-point function. The crucial point is that
subtracting a constant in momentum space corresponds to adding
a contact term proportional to $\delta^{(4)}(x)$ in position space. This contact term does not contribute to the Osterwalder--Schrader
condition for test functions supported in the positive-time half-space. Thus, reflection positivity is preserved even after renormalization. More generally, subtracting
a polynomial in $p$ corresponds to adding derivatives of delta
functions in position space; such contact terms are
also invisible to the Osterwalder–Schrader conditions. This is a consequence of
the principle that adding local contact terms does not change
the physical correlation functions away from coincident points. 

\section{Renormalized Källén-Lehmann representation}
\label{sec4}

Let us now show that the Euclidean first order two-point function integral \eqref{bubble}
also admits a Källén-Lehmann spectral representation, we need to
establish the following relationship
\begin{align}
    G_B(p^2)=\int_0^\infty ds \: \dfrac{\rho(s)}{p^2+s} \; .
    \label{K-L}
\end{align}
Here, $\rho(s)\geq0$ must be a positive spectral density function to
ensure the unitarity and causality of the theory. Is it clear
that if $G_B(p^2)$ diverges, just as
\eqref{bubble} does, then it is impossible to accomplish such
a relation directly. It is usual in such cases to use a suitably
subtracted spectral representation to deal with finite quantities. 
This subtracted/renormalized Källén-Lehmann representation
can be derived by taking the derivative of \eqref{K-L} with respect
to $p^2$ and then integrating from $p_0^2$ to $p^2$, i.e.,
\begin{align}
G_B^R(p^2)=G_B(p^2)-G_B(p_0^2)
\,=\,
\int_0^\infty ds \: \rho(s) \bigg(\dfrac{1}{p^2+s}
- \dfrac{1}{p_0^2+s}\bigg) \; ,
\label{renormalizedK-L}
\end{align}
where now $G_B^R(p^2)$ is finite. There are several ways of showing that $G_B^R(p^2)$ has a spectral
representation with positive spectral density. One way starts from the original first order two-point function expression
\eqref{bubble}. After analytic continuation  to Minkowski spacetime, we can use
Cutkosky cut rules. In this case, the result is almost immediate
\cite{Dudal2010wn}. The other way, which we detail here, is based
on a method  that was used in \cite{Baulieu:2009ha} for the case of
purely imaginary masses. Here, we will apply it to the more general case of complex masses. We start by considering the derivative of \eqref{bubble} with respect
to $p^2$, expressed in terms of Feynman parameters (the overall factor $1/(4\pi)^{2}$ is omitted for simplicity)
\begin{align}
\frac{d G_B(p^2)}{dp^2}
&=
- \int\limits^1_0 dx \,
\frac{x(1-x)}{x(1-x) p^2 + (1-x) m^2 + x m^{*2}}
\nonumber \\
&= - \int\limits^1_0 dx \, \bigg[
p^2 + \frac{(1-x) m^2 + x m^{*2}}{x(1-x)} \bigg]^{-1}
= - \int\limits^1_0 dx \, \bigg[p^2
- i \frac{(2x - 1) m^2_I + i m^2_R}{x(1-x)} \bigg]^{-1} ,
\label{int-der}
\end{align}
where we have defined $m^2=m_R^2+im_I^2$. Next, we perform the following change of variables:
\begin{equation}
\alpha = \frac{(2x - 1) m^2_I + i m^2_R}{x(1-x)}
= \begin{cases}
x = 0 &\rightarrow \;\;\; \alpha_0 = -(1 - i \, \theta_m)  \infty
\\
x = 1 &\rightarrow \;\;\; \alpha_1 = + (1 + i \, \theta_m) \infty
\end{cases} \hspace{0.5cm}
\text{with}\hspace{0.5cm}\theta_m = \frac{m^2_R}{m^2_I}\,.
\end{equation}
This means that the integration over $\alpha$ is along an infinite
line tilted at an angle $\theta_m$ with respect to the
real-$\alpha$ line. The integral in \eqref{int-der} then becomes
\begin{align}
\frac{d G_B(p^2)}{d p^2} &= - \int\limits^{\alpha_1}_{\alpha_0}
\left(\frac{dx}{d\alpha}\right) \frac{d\alpha}{p^2 - i \alpha}
\nn
 \\
&= - \,i \int\limits^{\alpha_1}_{\alpha_0}
\frac{d\alpha}{2 \alpha} \, \, \,
\frac{\alpha - 2m^2_I + \sqrt{\alpha^2 + 4 (m^2_I)^2
- 4 i m^2_R \alpha\;}}{(p^2 - i \alpha)^2}\,,
\label{int-alpha}
\end{align}
where the second term in the integrand is $x = x(\alpha)$. Note
that when solving $x$ in terms of $\alpha$, the root with the minus
sign in front of the square root is not the appropriate one, as it
does not yield the limiting values $x = 0$ and $x = 1$. We are working 
in Euclidean space $p^2 \ge 0$; therefore, the integral in~\eqref{int-alpha} 
does not have poles in the upper half-plane but has a square root branch cut 
starting from a threshold value $\alpha_{\rm thr}$. This threshold value is the
value of $\alpha$ for which the square root in the integrand
acquires an imaginary part; in the upper half plane, it is given by
\begin{equation}
\alpha_{\rm thr}
\,=\, i \left[ 2 m^2_R + 2 \sqrt{(m^2_R)^2+(m^2_I)^2} \right]  \; .
\label{alpha-thr}
\end{equation}
This means that one can transform (\ref{int-alpha}) into an integral
over a contour~$\Gamma$ in the upper half plane. The integral
over the contour at infinity vanishes because the integrand behaves
as $1/|\alpha|^2$ for large $\alpha$. Moreover, the integral over
the term $(\alpha - 2m^2_I)/2\alpha$ vanishes because of
Cauchy's theorem. Therefore, one is left with two integrals
(in opposite directions) along each side of the cut. %
To perform the integral over the square root more easily, we
change the variable again, $\alpha = i s$, so that
\begin{equation}
\alpha_{\rm thr} \,\,\longrightarrow \,\,
s_{0} = 2 m^2_R + 2 \sqrt{(m^2_R)^2 + (m^2_I)^2} \; ,
\label{s0}
\end{equation}
and
\begin{align}
\frac{d G_B(p^2)}{d p^2}
&= i \int_{\Gamma} ds \, \frac{1}{(p^2 + s)^2} \frac{\sqrt{- s^2 + 4 (m^2_I)^2 + 4 m^2_R s\;}}{2 s} \nonumber \\
& = i \int^\infty_{s_0} ds \, \frac{1}{(p^2 + s)^2}
\frac{i  \, \sqrt{s^2 - 4 (m^2_I)^2 - 4 m^2_R s\;}}{2 s}
\nonumber
\\
&+i \int^{s_0}_\infty ds \, \frac{1}{(p^2 + s)^2}
\frac{-i  \, \sqrt{s^2 - 4 (m^2_I)^2 - 4 m^2_R s\;}}{2 s}
\nonumber \\
& = - \int^\infty_{s_0} \,
\frac{ds }{s\,(p^2 + s)^2} \;\sqrt{s^2 - 4 (m^2_I)^2 - 4 m^2_R s} \,\,. \end{align}
This last expression confirms the well-known result that the integral of
a function with a branch cut is twice the imaginary part of that
function. Now, integrating over $p^2$ and recovering the overall factor, we obtain
\begin{equation}
G_B^R(p^2) = G_B(p^2) - G_B(p_0^2)
\,=\,
\int^\infty_{s_0} ds \,
\frac{\sqrt{s^2 - 4 (m^2_I)^2 - 4 m^2_R s\;}}{(4\pi)^2s}
\left(\frac{1}{p^2 + s} -\frac{1}{p_0^2+s}\right) \; ,
\label{F-fin}
\end{equation}
where $G_B(p_0^2)$ is a finite real number corresponding to the subtraction, 
which is crucial for taming the UV divergence. Is it clear now, 
comparing~\eqref{F-fin} and~\eqref{renormalizedK-L}, that the first order two-point 
function integral~\eqref{bubble} satisfies a renormalized Källén-Lehmann representation 
with a manifestly non-negative spectral density given by
\begin{equation}
\rho(s)\,=\,\Theta(s-s_0)\frac{\sqrt{s^2 - 4 (m^2_I)^2
- 4 m^2_R s\;}}{(4\pi)^2s} \,\geq \,0 \; .
\end{equation}
In Minkowski space, this corresponds, up to a $\pi$ factor, to
the discontinuity of the first order two-point function $G_B(p^2=-\tau+i\epsilon)$ along its
branch cut, where it develops an imaginary part. An easy proof of
this general statement can be seen directly from the Källén-Lehmann
representation \eqref{K-L} of a correlator $F(p^2)$ and by using the
Sokhotski–Plemelj identity:
\beq
&&
\lim_{\,\ep \rightarrow 0^+} \text{Im}[F(p^2=-\tau-i\epsilon)]
\,=\lim_{\epsilon \rightarrow 0^+}\dfrac{1}{2i}
\Big[
F(-\tau-i\epsilon)-F(-\tau+i\epsilon)\Big]
\nonumber
\\
&&
\qquad
\qquad
\qquad
\qquad
\qquad
\qquad
=\,\,
\lim_{\epsilon \rightarrow 0^+} \dfrac{1}{2i}
\int_0^\infty ds \:\rho(s)\left( \dfrac{1}{-\tau-i\epsilon+s}-\dfrac{1}{-\tau+i\epsilon+s}
 \right) \nonumber
\nonumber
\\
&&
\qquad
\qquad
\qquad
\qquad
\qquad
\qquad
=\,\,
\dfrac{1}{2i}\int_0^\infty ds \:\rho(s)\left(2\pi i\delta(\tau-s) \right)
\,=\, \,\pi\rho(\tau)\; .
\label{imaginary}
\eeq

Once a renormalized Källén-Lehmann representation is obtained for the renormalized first order two-point function $G_B^R(p^2)$, it is straightforward to obtain such a representation
for the resumed two-point correlation function $C^R(p^2)$ defined in
\eqref{GB-pole} with the first order renormalized function $G_B(p^2)$. Using the same method that was
used to derive the Källén-Lehmann decomposition of
$G_B(p^2)$, we can obtain the spectral function for $C^R(p^2)$
by computing its imaginary part in Minkowski space, as in
\eqref{imaginary}. From the definition of the correlator, we have
\beq
&&
\lim_{\epsilon\rightarrow0^+}\text{Im}
\big[C^R(p^2=-\tau-i\epsilon)\big]
\,\,=\,\,\lim_{\epsilon\rightarrow0^+}
\text{Im}\left[\dfrac{G_B^R(-\tau+i\epsilon)}{1
-\lambda_{12} G_B^R(-\tau+i\epsilon)}\right]
\nonumber
\\
&&
\qquad
\qquad
\qquad
=\,\,\dfrac{\text{Im}[G_B^R(-\tau)]}{(1-\lambda_{12}\text{Re}
[G_B^R(-\tau)])^2+(\lambda_{12}\text{Im}[G_B^R(-\tau)])^2}
\,=\,\pi \rho_C(\tau) \; , \label{rhoC}
\eeq
where $\rho_C(s)$ corresponds to the spectral density of the continuum part (above the branch cut) of the resumed correlator $C^R(p^2)$. An explicit expression for $\rho_C(\tau)$ above \eqref{rhoC} is obtained from the real and imaginary parts of the renormalized expression in Minkowski space, derived by analytic continuation of the first order correlation function, where we have fixed $m^2=1+i$ and $p_0=1$ for concreteness:
\beq
&&
\text{Re}\,
\left[G_B^R(-\tau)\right]
\,=\,
\text{Re}
\Bigg\{
-\frac{1}{32\pi^2\,\tau}\Bigg(
\pi(-\tau - 1) - \tau\log 2
\nonumber
\\
&&
\quad
+ \,4\sqrt{(4-\tau)\tau + 4}
\bigg[
\tan^{-1}\!\bigg(
\frac{-\tau + 2\sqrt{2} + 2}{
\sqrt{(4-\tau)\tau + 4}}
\bigg)
\,-\,
\tan^{-1}\!\bigg(
\frac{2 - \tau}{\sqrt{(4-\tau)\tau + 4}}\bigg)
\bigg] \Bigg)\Bigg\},
\qquad
\eeq
where
\beq
\text{Im}[G_B^R(-\tau)]
&= \pi\Theta(s-(2+2\sqrt{2}))\frac{  \sqrt{\tau ^2-4-4\tau}}{(4 \pi )^2 \tau } \, . 
\eeq

Of course, this is not the whole spectral density for the correlator, since 
by the definition of $C^R(p^2)$ we also have an isolated contribution from the 
pole associated with the ghost condensate bound state of mass $\mathcal M$. 
The complete renormalized Källén-Lehmann representation of the resumed two-point
correlation function is then
\begin{align}
C^R(p^2) \,\,=\,\,
\int^\infty_0 ds \;\rho(s) \left(\frac{1}{p^2 + s}-\frac{1}{p_0^2 + s}\right)
\; ,
\end{align}
with
\begin{equation}
\rho(s) = R_C\, \delta(s - \mathcal{M}^2) + \rho_C(s) \, \Theta(s - s_0) \ge 0 \; , \;\;\;\;\; R_C=-\frac{1}{\lambda_{12}^2\text{Re}[\frac{d}{ds}G_B(-s)\vert_{s=\mathcal M^2}]}\; . \label{rho-KLR}
\end{equation}
Here $R_C$ is the residue corresponding to the bound state, and
$s_0$ is the threshold of a physical branch cut. Note that for the
noninteracting theory, $\lambda_{12} = 0$, we obtain the spectral
function of the first order two-point function, as it should. Also, the value of
$s_0$, given by \eqref{s0}, can be rewritten as
\beq
&&
s_0 \,\,= \,\,2 m^2_R + 2 \sqrt{(m^2_R)^2+(m^2_I)^2}
\,\,=\,\, m^2 + m^{*2} + 2 \sqrt{m^2  m^{*2}}
\nn
\\
&&
\qquad \qquad
=\,\, \left( \sqrt{m^2}
+ \sqrt{m^{*2}} \right)^2 = \left(m + m^*\right)^2 \; ,
\eeq
which generalizes the well-known result of the branch cut for 
real masses. Using the results of the first section, as we have 
obtained a renormalized Källén-Lehmann representation for the 
resumed two-point function correlator, we can claim that it also 
satisfies reflection positivity, since only a momentum independent 
subtraction (contact term) is needed to handle the UV divergence. 

\section{Conclusions}
\label{secConc}
We have investigated the consistency of a six-derivative 
scalar field theory featuring a bound state formed by a pair of ghost 
fields with complex-conjugate masses that we have proposed in~\cite{AKPS-bounds}. 
We examined whether this model satisfies two fundamental consistency conditions 
for a quantum field theory: 
Osterwalder–Schrader reflection positivity and the existence of 
a renormalized Källén–Lehmann spectral representation with a positive 
spectral density. We have shown that both conditions are satisfied by 
the two-point function of the composite (bound state) operator, 
indicating that physical observables can consistently emerge from the
underlying ghost dynamics in such a theory.
In fact a quantum field theory is not completely defined until one specifies which are their physical fields. In the present case they are the composite local fields 
\beq \nonumber
O_{\varphi_1\varphi_2}(x) = \varphi_1(x) \varphi_2(x) \ , 
\eeq

 Our main goal is to explore the possibility of forming bound 
states in higher (six or more) derivative models of quantum gravity,
where  ghost-like states with complex-mass poles naturally appear. For this reason, 
it would be interesting to extend the simplest toy model  Ref.~\cite{AKPS-bounds} endowed with a mechanism of confining  ghosts into bound states  by including more general types of interactions, e.g., 
those with derivatives, which are typical for quantum gravity models. 
One can foresee that the methods of proving the consistency of
the theory with bound states, which we used here for the relatively
simple scalar model with the single quartic interaction term
(\ref{U12}), may eventually be used for more sophisticated models of quantum gravity involving ghosts in a controlled way.

\section*{Acknowledgements} 

G.K. acknowledges discussions with David Dudal on spectral representations of 
bound states of particles with complex masses. This work was supported by Spanish 
Grants no. PGC2022-126078NB-C21 funded by MCIN/AEI/10.13039/501100011033, 
Diputación General de Aragón-Fondo Social Europeo (DGA-FSE) Grant no. 2020-E21-17R 
of the Aragon Government; and the European Union, NextGenerationEU Recovery 
and Resilience Program on Astrofísica y Física de Altas Energías, 
CEFCA-CAPA-ITAINNOVA (M.A.); Conselho Nacional de Desenvolvimento Científico 
e Tecnológico (CNPq) grants nos. 305122/2023-1 (I.Sh.), 313254/2025-7 (G.K.); 
Fundação de Amparo à Pesquisa do Estado de São Paulo (FAPESP), grant 
no. 2018/25225-9 (G.K.).

\end{document}